\documentclass{article}

\usepackage{arxiv}
\usepackage{eurosym}
\usepackage[utf8]{inputenc} 
\usepackage[T1]{fontenc}    
\usepackage{hyperref}       
\usepackage{url}            
\usepackage{booktabs}       
\usepackage{amsfonts}       
\usepackage{nicefrac}       
\usepackage{microtype}      
\usepackage{lipsum}
\usepackage{graphicx}
\usepackage{color}
\usepackage{amssymb}
\usepackage{subcaption}
\usepackage{amsmath}
\usepackage[autostyle]{csquotes}
\usepackage{algorithm}
\usepackage{algorithmic}
\usepackage{textcomp}
\usepackage{trimclip}
\usepackage[algo2e]{algorithm2e}

\title{Reshaping consumption habits by exploiting energy-related micro-moment recommendations: A case study}

\author{Christos Sardianos \thanks{This paper will appear in Communications in Computer and Information Science( CCIS) - Springer Book - [Smartgreens extension]}, Iraklis Varlamis, Christos Chronis, George Dimitrakopoulos \\
Department of Informatics and Telematics\\
Harokopio University of Athens\\
Athens, Greece\\
\texttt{sardianos@hua.gr;varlamis@hua.gr;it21797@hua.gr;gdimitra@hua.gr} \\
  \And
Abdullah Alsalemi, Yassine Himeur, Faycal Bensaali\\
  Department of Electrical Engineering\\
  Qatar University\\
  Doha, Qatar \\
  \texttt{yassine.himeur@qu.edu.qa;a.alsalemi@qu.edu.qa;f.bensaali@qu.edu.qa} \\
   \And
 Abbes Amira \\
  Institute of Artificial Intelligence\\
  De Montfort University\\
  Leicester, United Kingdom \\
  \texttt{abbes.amira@dmu.ac.uk} \\
}

\begin{document}
\maketitle

\begin{abstract}
The environmental change and its effects, caused by human influences and natural ecological processes over the last decade, prove that it is now more prudent than ever to transition to more sustainable models of energy consumption behaviors. User energy consumption is inductively derived from the time-to-time standards of living that shape the user's everyday consumption habits. This work builds on the detection of repeated usage consumption patterns from consumption logs. It presents the structure and operation of an energy consumption reduction system, which employs a set of sensors, smart-meters and actuators in an office environment and targets specific user habits. Using our previous research findings on the value of energy-related micro-moment recommendations, the implemented system is an integrated solution that avoids unnecessary energy consumption.  With the use of a messaging API, the system recommends to the user the proper energy saving action at the right moment and gradually shapes user's habits. The solution has been implemented on the Home Assistant open source platform, which allows the definition of automations for controlling the office equipment. Experimental evaluation with several scenarios shows that the system manages first to reduce energy consumption, and second, to trigger users' actions that could potentially urge them to more sustainable energy consumption habits.
\end{abstract}


\section{Introduction}
\label{sec:introduction}
The rise of living standards in modern society over the last years has led to a surge in the daily use of technology devices and appliances \cite{hu2017survey}, which in turn increased the consumption of energy resources and gave rise to new environmental and socio-economic problems \cite{URGEVORSATZ2013141}. As a counterpart, technology plays an assisting role in helping users improve their energy efficiency levels. However, most of smart-home and energy related automation systems focus on increasing user's ease of access in controlling or monitoring household appliances \cite{jensen2018designing,darby2018smart}, but still, the choice of managing the use of these appliances solely relies on the environmental and economical awareness of the user.

Despite the fact that technology provides the means for efficient energy consumption, it is the user behavior that plays the most important role in forming the household's energy footprint~\cite{gram2013efficient}. Hence, it is important to motivate users ---who are not committed and self-motivated--- and to increase the awareness about contemporary energy issues and their dramatic repercussions. This is a key factor for increasing individual energy efficiency and consequently reducing the energy footprint of a community.

Considering the impact of motivating the user to change their everyday energy consumption, we identify the need for information technology solutions that address the problem of engaging users in adopting more sustainable energy usage tactics \cite{coutaz2018will}. Everyday energy-related behavior is definitely driven by the user needs and desires. However the behavior is synthesized by many small actions, which are influenced by external factors, such as outdoor temperature and humidity (e.g. turning air-conditioning on when it is hot) and by the user's common habits (e.g. switching the water heater on after arriving home to take a bath). In tandem, user needs, user conditions and user habits shape the user's energy consumption profile.

Recommender systems aim at providing data-driven recommendations to users\cite{ricci2011introduction}. In the case of raising energy awareness and changing users' energy habits, these systems can be used to recommend energy-related actions to the users that could potentially affect their consumption footprint. But first, they must be able to identify users' behavior \cite{zhou2016understanding}
in order to provide recommendations that match user profile and have a high potential of being accepted. Then, they must be capable of recommending the correct action to the user, at the right moment in order to maximize the acceptance probability. Finally, they must adjust to the user needs and gradually modify the user habits. Such personalized recommendations are also most likely to be adopted by the user in the long term and gradually transform user behavior towards energy efficiency. In order to maximise the benefit from habit change, recommendations must target frequently repeated actions and also actions that maximise the reduction of the energy footprint. They must take into account the environmental conditions (e.g. weather) and avoid inelastic user activities that have a specific appliance usage duration (e.g. cooking). As a result, any attempt to reduce energy footprint based on user habit must follow an analysis of the user habits and must be based on specific usage scenarios.

As mentioned earlier, actions that relate to energy consumption may differ among users depending on their habits, but also can be affected by external conditions (e.g. weather and season changes) or individual user needs, which both may change over time.
Considering the repetitive nature of user habits and the temporal change in user needs and external conditions, it is necessary that any predictions or recommendations about user near-future actions must combine both types of information in order to improve efficiency. In this direction, we examine user's daily activities in segments, which are called user ``\textit{micro-moments}", in adoption of the term introduced by Google \cite{em3_ramaswamy2015micro} for capturing the temporal nature of smartphone usage for covering information needs.

The proven success of micro-moments in information search and retrieval \cite{snegirjova2017micro} can be adopted by personalized assistants that analyze contextual information from various sources (e.g. GPS data, environmental information, user status and mood, etc.), predict user needs and pro-actively recommend pieces of information or activities to the user that may be useful at that specific moment \cite{campos2014time} or place \cite{bao2012location}.

In our previous work \cite{sardianos2019Iwanttochange}, we demonstrated how the analysis of energy consumption logs can highlight repetitive usage patterns in home appliances. These patterns correspond to the usage of specific devices at specific time-slots, every weekday or weekend and have been repeated during the whole monitoring period, thus allowing recommendations to be addressed at the right micro-moment (e.g. a few minutes before the typical usage duration ends)\cite{alsalemi2019role}. In this work, we build on this concept and present the system that we developed in order to take advantage of micro-moment based recommendations. The recommendations target specific usage scenarios in an office context and the whole system is implemented on a popular open source platform for home monitoring using a set of sensors and actuators and a set of automations that help reducing unnecessary energy consumption.

The main contributions of this work can be summarized as follows:
\begin{itemize}
    \item A methodology for the analysis of energy consumption data and the extraction of habits, which has been applied on a real dataset.
    \item A system implementation that integrates sensors, actuators, automations and a messaging mechanism for recommending energy saving actions to the user at the right moment.
    \item An experimental evaluation in a real office setup of the system. The setup  controls the light and monitors. Unnecessary or excessive usage is avoided by notifying the user to turn off the devices when necessary.
\end{itemize}

In section \ref{sec:relwork}, we summarize the most important works on recommendation systems for energy efficiency and discuss the concept of micro-moments. Section \ref{sec:methodology}, starts with a motivating example and then provides an overview of the proposed methodology, which has been extensively presented in a previous work \cite{sardianos2019Iwanttochange}. In section \ref{sec:implementation}, we provide the details of the implemented system architecture and in section \ref{sec:results} we present the implementation results. Finally, section \ref{sec:conclusions} summarizes the progress so far and the next steps of this work, which is expected to deliver a fully operational system that monitors user energy consumption habits and timely recommends action that can reduce the energy footprint.

\section{Related Work}
\label{sec:relwork}
The concept of mining useful knowledge from usage logs has been discussed several times in the related literature. Although the initial focus back in 2000 was in web browsing and web usage logs \cite{srivastava2000web}, there are several recent works that mine user activity logs, outside of the web browsing environment, including geo-location logs \cite{sardianos2018extracting}, app usage logs \cite{cao2017mining}, bio-signal logs \cite{alhamid2013leveraging}, etc.
The aim of geo-location log mining works is to discover hidden patterns in the user's daily behavior and either highlight interesting locations and travel sequences \cite{cao2010mining} or create recommendations for Location-Based Social Networks \cite{bao2015recommendations}. Overall log mining approaches, analyze the activity logs of many users in order to detect the common context in which certain activities are preferred among users. Consequently, these patterns and the user’s personal context-aware preferences are utilized in order to create personalized and context-aware recommendations \cite{yu2012towards}.

The term ``micro-moments" has been introduced in the literature with the `Janus Factor' theory for determining marketing behavior \cite{em3_stokes2012micro} and describe the moments where people are positively positioned towards buying something promoted by a campaign and moments where people are skeptical and difficult to persuade.
Google coined the concept of micro-moments to the spontaneous interaction with smartphones in order to learn, discover, carry out an activity, or buy  a product online \cite{em3_ramaswamy2015micro}, but it soon has been expanded to more fields, introducing new types of micro-moments that span daily life and can be appropriate for the tourism industry (e.g. I want to show \cite{em3_jorgensen2017want}, I want to remember \cite{em3_wang2012role,em3_bilovs2016micro}).

Loviscach \cite{loviscach2011design} proposed a conceptual computer-based energy efficiency recommendation system for domestic households. The solution combines the use of digital power meters, heating control systems, smartphone apps. Also, it suggests making use of deliberate decisions through adopting machine earning algorithms in order to conduct automated energy saving is also studied. The paper acts as a blueprint for future work to build upon.

In \cite{Ovidiumulti}, a surrounding smart system that supply step-by-step recommendations at different stages such as; behavior, appliance-level and consumer-feedback is proposed. The main focus of this work is the design of a recommendation architecture for improving energy consumption behavior especially in households and dwelling areas.

In \cite{Schweizer-using}, energy efficiency is achieved through studying consumption profiles and preferences of the occupants. This was conducted from one side; by using a sequential pattern mining system adapted to meaningful household power usage data, and from another side by introducing a recommender system which delivers advices to consumers to decrease their power usage. Moreover, this system was tested on several households where a group of participants is selected to make rating on the influence of the recommendations on their well-being. This rating was finally deployed to adapt the system variables and turn it more precise over a second test step.

In \cite{Hiroki-human}, a recommender system, namely, KNOTES is proposed. It represents a social experience on a consulting recommender system for household power-saving. KNOTES attempts to generate powerful suggestions through the consumer’s own preferences based on its value feeling and relevance of the guidance. Further, a recommendation reference codebook is employed, which helps avoiding duplication.

In \cite{Starke:2017:EUI:3109859.3109902}, two frameworks are presented for supporting consumers making power consumption behavioral change through drawing tailored energy efficiency guidance. Both architectures employ ordinal Rasch scale which uses 79 behavioral power-efficiency rules and connect this to a consumer power efficiency aptitude for tailored guidance. Consequently, making use of of Rasch-based guidance helps optimizing consumer’s effort, raises choice gratification and conducts to the endorsement of more power efficiency actions.

In \cite{zanbouri-data}, the matrix factorization approach is deployed to develop a recommender system for advising energy-saving e-service. The authors advocated to practice database features as a contextual data which could be incorporated in the e-service based recommender system since most of related works did not consider data characteristics in developing e-service recommender systems.

In \cite{Jimenez-multi}, a recommender system is built relying on a multi-agent system allowing the collection of power consumption patterns from sub-meters deployed in a household, gathering online electricity bills and generating advices based on usage profiles and consumer energy cost. By doing so, the recommender system suggests new moments for using the electric equipment and devices with lower price, bringing then, the cost-effectiveness for the consumers. In this sense, optimizing power usage and reducing peaks and outages is potentially attained.

In \cite{kar2019revicee}, the authors proposed a simple yet effective recommendation system that learns personal and collaborative consumer-preferences from archived energy usage information and then affords advices for smart home lighting controls. To this end, the smart part is accomplished through deriving a group of points for monitoring light actions when keeping a good balance between individual visual comfort and power efficiency.

In \cite{luo2017non}, a personalized recommendation system is proposed that utilizes recommendations for deducing consumer’s possible concerns and requirements for power efficient devices, then recommending power efficient appliances to consumers, and accordingly engendering possibilities to preserve energy in distribution network. The recommender system is developed through carrying out a non-intrusive appliance load monitoring (NILM) scheme. It operates generalized particle filters to separate the consumer’ house device usage patterns from the smart-metering records. Using NLIM outputs, a set of inference tasks was exercised to predict preferences and power usage outlines. Finally, the device profile is derived using information retrieval algorithms extracting key-words from textual device advertisements.

The current work, builds on a new type of micro-moment, which is related to the behavioral change of users towards energy efficiency. We call this the ``I want to change'' moment. Such moments are used to deliver the correct recommendation to the user to assist him/her to adopt a better behavior. In order to gradually achieve this habitual behavior change towards energy-efficiency, we must first detect the micro-moments by analyzing user contextual logs, associate micro-moments with specific user activities and recommend actions that can assist the user to reduce his/her energy footprint. In the following section we give a motivating example and then describe the proposed methodology.

\section{Methodology}
\label{sec:methodology}

Our previous work \cite{sardianos2019Iwanttochange} was motivated by the need to provide users with the means to improve their energy consumption profile. So it introduced a context-aware recommendation system that analyses user activities and extracts their habits. Based on these habits, the system is able to present personalized energy efficiency recommendations at the right micro-moment and place.

In order to transform users' energy habits, it is important first to detect them by processing their activity logs and then to provide the appropriate motivations that will help them change. According to the ``habit loop'' theory \cite{em3_duhigg2013power} a typical habitual behavior goes through three stages: i) the \textit{cue}, a trigger that puts the brain to auto-pilot, ii) the \textit{routine} that refers to the actual action performed by the individual following the cue and iii) the \textit{reward}, which is the satisfaction induced from completing the routine and an indicator to the potential to repeat the behavior. Reconstructing a bad habit loop into a better one requires detecting the cue, modifying the routine and demonstrating the reward in order to strengthen the desired habitual behavior.

\begin{figure}[h!]
\includegraphics[width=0.75\columnwidth]{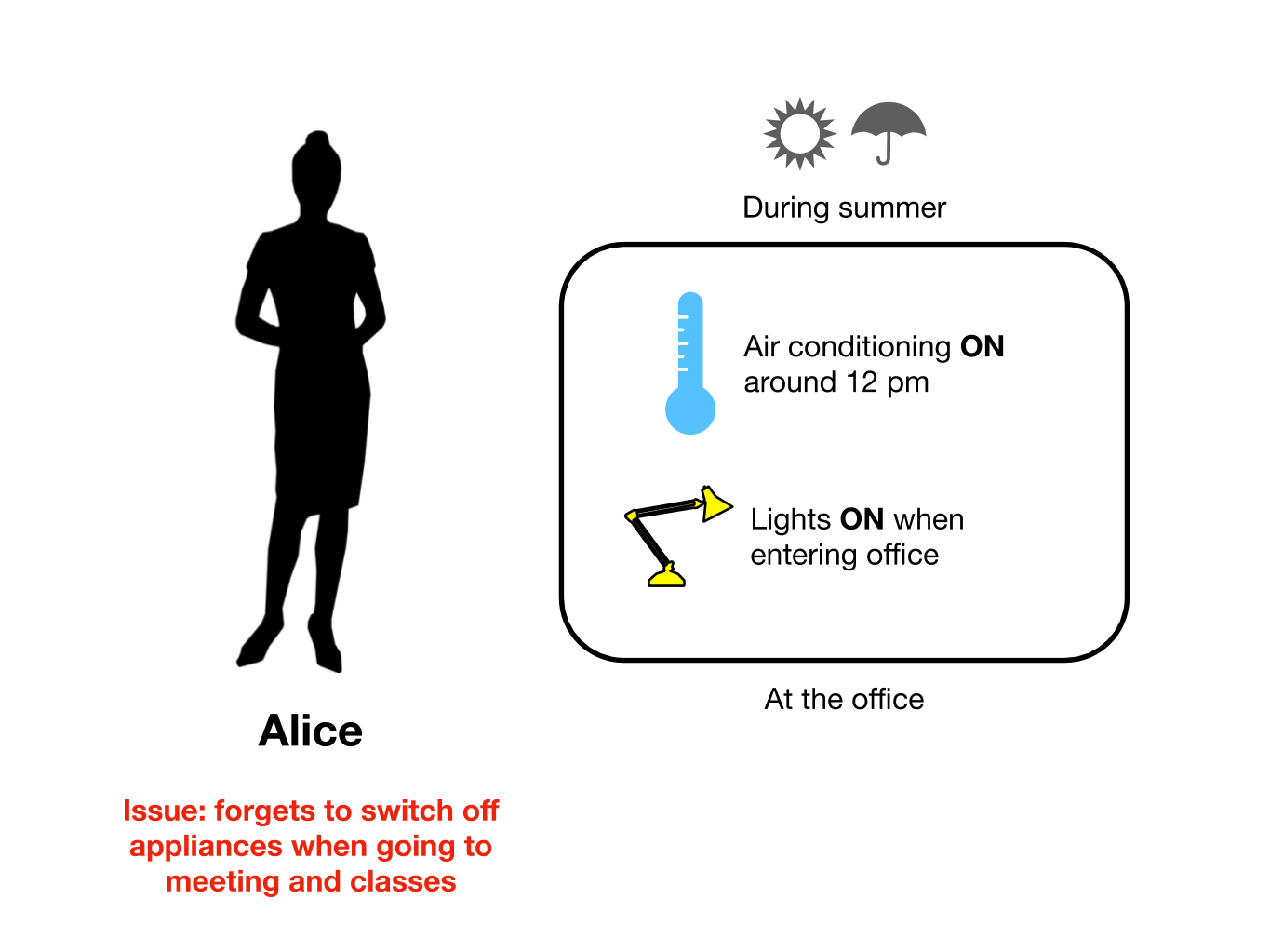}
\centering
\caption{Alice's Use Case}
\label{fig:em3-alice}
\end{figure}

In order to better understand the role of micro-moment based recommendations, we may consider the case of a user, Alice, at her  office at the university campus, as depicted in Figure \ref{fig:em3-alice}. \textit{``Alice usually arrives at the office around 8 o'clock and starts working at the office, until she has to go to classes. During classes she is away from the office for three or more hours and she usually forgets to turn off the office lights and the computer monitors. When she returns from class, she usually has meetings with students and colleagues so she may be absent from her office for more time. In some cases, despite the outdoor conditions that may offer sufficient natural lighting, she keeps having the office lights on during the whole day. In addition, Alice tends to turn the air-conditioning system on when she arrives at the office, especially at the hot summer days, but she forgets to turn it off when leaving the office. There are also cases that she used to let the air-condition on even after returning back home at the evening in order to keep the temperature leveled until she returns the next morning.''}

The proposed recommendation framework that we have described in \cite{sardianos2019Iwanttochange} analyzes historical information about the user's daily consumption and correspondingly extracts consumption habits. The habits result from a generalization of user activities in time and external conditions. In our example, the user habit concerning 
the usage of office lights and monitors will be as follows: 
\textit{``The lights are switched on between 9am and 5pm during weekdays, the monitors are also on for the same periods, unless Alice switches them off. They fall into standby mode after 30 minutes of computer inactivity. Usually, Alice forgets to switch off the lights when she leaves the office in the afternoon, resulting raised cumulative expenses over the year"}.

In the example above, the exact times when the on and off activities happen are user-dependent (i.e. if it is at 1:00 pm or at 1:05 pm, etc.) but definitely link to Alice's schedule (e.g. before or after her weekly classes). The same holds for the exact time when she arrives and leaves the office.  Similar scenarios can be applied to the usage of the air-cooling device, but they also depend on the weather conditions. Based on the actual weather and light conditions and the actual user status (e.g. user entered or exited the office) a recommendation for a repetitive action that has been properly positioned within the user daily schedule, or has been smartly shifted a few minutes earlier or later in order to save energy, will be more than welcome for the user. Such a recommendation will increase the user's trust in the recommendation system and will assist her not only to reduce expenses, but also to boost the overall sustainability footprint of the organization.

\begin{figure*}[h!]
\includegraphics[width=0.8\columnwidth]{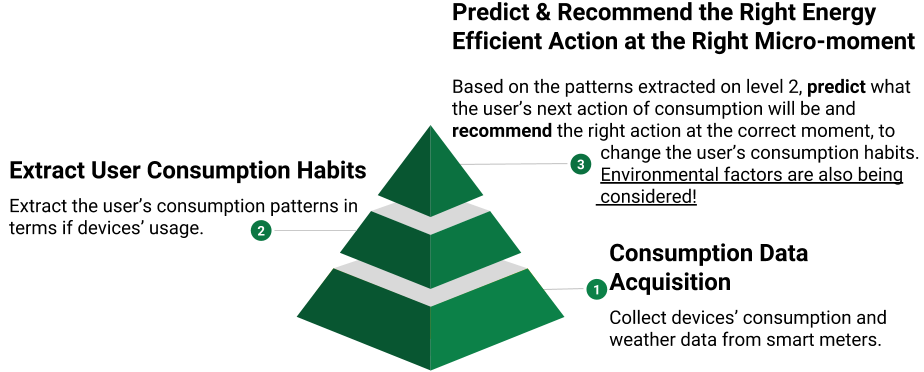}
\centering
\caption{Steps for data analysis and creating energy consumption-related recommendations. (Also see \cite{sardianos2019Iwanttochange}, Section 3)}
\label{fig:steps-pyramid}
\end{figure*}

The process of creating energy efficient recommendations is based on a three-step approach as depicted in Figure \ref{fig:steps-pyramid}. The first step of the process refers to the consumption data acquisition and analysis. Based on the analysis of the user's consumption data along with environmental conditions we perform an initial step of analysis to extract meaningful insights. So we process the Consumption Logs and Weather Logs to highlight the user’s consumption actions in terms of  micro-moments and extract the context of user activities (i.e. when the user tends to switch on and off a specific device). Being able to identify user's energy demands on the spot, in the third step of our approach we can predict user's next energy consumption activities (e.g. 10 minutes after entering the office the user switches on the air-conditioning, when the natural light is above a threshold, the user does not turn on the lights etc.), which enables our recommender system to recommend energy-related actions to the user beforehand, so as to lead his energy profile in higher levels of efficiency. In the sections that follow we detail on each step of the process.

\subsection{Data Acquisition and Analysis}
\label{data-acquisition}

In order to collect the user's consumption data, we rely on WiFi-enabled smart plugs/outlets equipped in the most frequently used appliances (e.g. Sonoff Pow R2) \cite{em3_sonoff_pow}. These smart outlets provide information about the energy consumption of each device, which we collect in a minute window. More specifically, all user's devices that consume big loads of energy are plugged in a smart plug that measures the consumption in KWh in real time. In addition, a variety of sensing modules are also installed in the place for recording contextual information (e.g. temperature, humidity, luminosity and occupancy).

\subsection{Data Transformation and Preprocessing}

Data collection is the first step of the process, but in order to exploit the above types of information, it is necessary to perform an initial step of analysis that will produce meaningful insights for the next steps of the pipeline. 
Consumption data are processed and a first level of abstraction is performed that maps on/off actions to device usage periods and characterises usage as normal, excessive, etc. In a recent work \cite{alsalemi2019endorsing}, we demonstrate how we characterised device usage with the help of consumption data and a classification algorithm. This analysis highlights the user's consumption actions in terms of micro-moments and then with the analysis of sensor data it is possible to extract information about the context of each user action.

As a proof of concept, in the previous work \cite{sardianos2019Iwanttochange}, we examined an online dataset provided by the University of California Irvine through its machine learning dataset repository. The dataset\footnote{\url{ https://archive.ics.uci.edu/ml/datasets/individual+household+electric+power+consumption}} concerns the monitoring of individual household electric power consumption. It is a multivariate time-series dataset with 2,075,259 measurements gathered in a house located in Sceaux (7km of Paris, France) between December 2006 and November 2010 (a time period of 47 months). The dataset contains information about the household global minute-averaged active power (in kilowatt), household global minute-averaged reactive power (in kilowatt) and minute-averaged voltage (in volt). The measurements concern the energy metering (in watt-hour of active energy) of three rooms of the household, the kitchen (which contains mainly a dishwasher, an oven and a microwave), the laundry room (which contains a washing-machine, a tumble-drier, a refrigerator and a light) and a set of energy consuming devices which correspond to an electric water-heater and an air-conditioner. 

In the pre-processing step (see Algorithm \ref{alg1}), we applied a time-series analysis methodology on the consumption information data of each room. More specifically, from the actual energy consumption recorded per minute, we computed the changes between consecutive minutes and between consecutive 5-minutes periods. The first feature allowed us to isolate minutes where the power consumption increased or decreased significantly due to powering on or off one or more devices. By applying a k-Means clustering on the different power change values recorded for a room, we obtained a number of clusters of power change values that we mapped to specific actions of operating multiple devices.

The processing of the consumption log file determined when a device has been turned on or off with ample accuracy. Subsequently, it allowed the extraction of the user's actions along with the moment that they took place in terms of micro-moments (e.g. at 10:34:00 AM (GMT) the user turned on the microwave and at 11:21:00 AM (GMT) turned on the dishwasher).
In order to map power changes to user actions we assumed that each device has a typical consumption specification. For this purpose, we adopted the values provided by the `energy calculator' website\footnote{\url{https://www.energyusecalculator.com/calculate_electrical_usage.htm}} in order to get an estimation of the devices' monitored power consumption. Based on the consumption values of each device, and the devices per room, we map power changes to user actions in terms of micro-moments.

\begin{algorithm}
\caption{Characterize device operation action as \textit{on} or \textit{off}.}
\begin{algorithmic} 
\REQUIRE Series of consumption data of each room recorded per minute.

\STATE $CurrentMinute = 0$
\LOOP
    \item Detect significant power consumption changes.
    \IF{$CurrentMinute = 5$}
        \item Detect significant power consumption changes in 5\-minutes periods.
        \item $CurrentMinute = 0$
    \ENDIF
    \item $CurrentMinute \leftarrow CurrentMinute + 1$
    
\ENDLOOP
\STATE Apply k-means algorithm on the power change values.
\STATE Find clusters of power change values.
\STATE Find the limit values between power changes.
\STATE Map changes to actions of operating devices.
\end{algorithmic}
\label{alg1}
\end{algorithm}

\subsection{Detection of user habits}
After this process, user's micro-moment data are analyzed to extract frequent itemsets (i.e. user action and contextual conditions sets).
The identification of user's frequent energy consumption habits is a key factor in the whole analysis, since this enables our system to create personalized energy saving action recommendations that shape user's consumption habits and impact the total energy consumption profile.

The habit detection task, refers to the process of identifying frequent consumption patterns (or device usage patterns) in the consumption micro-moments and associating them with weather conditions and other temporal parameters (e.g. time of day, day of the week etc.).
In this step, an Association Rule Mining algorithm is employed in order to jointly process user's micro-moments data (consumption and weather conditions) and find frequent itemsets (condition sets) that are associated with an action. 
The Apriori association rule extraction algorithm \cite{agrawal1994fast} is used to uncover how items are associated to each other by locating frequently co-occurring items among the users' transactions.

The typical example for describing association rule discovery algorithms is with the analysis of user shopping carts in an online shop. Let \(I = \{i_1, i_2, ..., i_n\}\) be all the possible itmes  that can be found in a cart and \(D = \{t_1, t_2, ..., t_n\}\) be the set of all transactions (shopping carts) in the shop's database. Each transaction in \(D\) contains a subset of the items in \(I\). If \(X, \Upsilon \subseteq I\) and \(\cap = \emptyset\), then the rule \(X \to \Upsilon\) implies the co-occurrence of \(X\) and \(\Upsilon\), meaning that if item \(X\) is bought, then item \(\Upsilon\) will also be bought together. By definition\footnote{\url{http://software.ucv.ro/~cmihaescu/ro/teaching/AIR/docs/Lab8-Apriori.pdf}} ``\textit{the sets of items (for short itemsets) \(X\) and \(\Upsilon\) are called antecedent (left-hand-side or LHS) and consequent (right-hand-side or RHS) of the rule respectively.}''

In the energy recommendations scenario, the appliance, the space, the time and the other conditions are the items that can be found in the LHS part of an association rule and user actions (i.e. switch the appliance on or off) are the consequent (RHS of the rule.).

\subsection{Usage analytics, recommendations and user feedback}
\begin{figure}[h!]
\includegraphics[width=0.85\columnwidth]{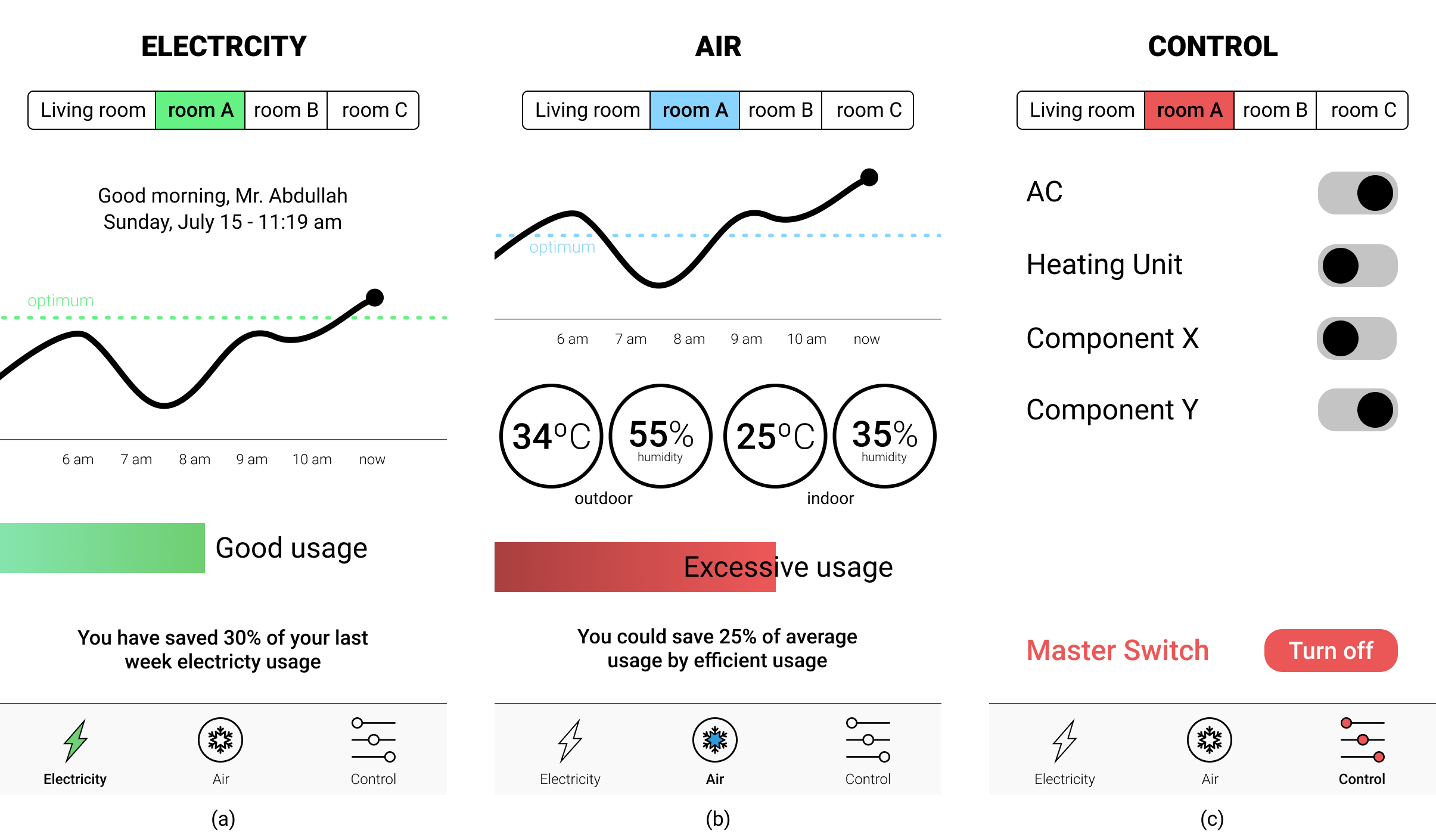}
\centering
\caption{(EM)\textsuperscript{3} Energy Tracking Application. (Also see \cite{sardianos2019Iwanttochange}, Section 3)}
\label{fig:em3-mobile-app}
\end{figure}

In order to demonstrate the effect of this change of energy consumption habits, the system collects and analyses user's consumption data and provides useful analytics to the user. For this purpose, the system can use a mobile app (such as the (EM)\textsuperscript{3} Energy Tracking Application which we have developed and is shown in Figure \ref{fig:em3-mobile-app}) or any other desktop application (as shown in the following section).

\section{Energy consumption monitoring system
\label{sec:implementation}}

\subsection{Case description
As explained in section \ref{sec:methodology}, the target of the demonstration scenario is a university employee, who spends a lot of time away from the office during the day, but still leaves the lights, monitor and A/C on.}

The ultimate goal is to assist the user to reduce excessive usage of lights, monitors' usage and A/C power by consorting with her usage habits. 
For this purpose, we implement a monitoring system that utilizes smart meters and sensors to monitor device consumption and Alice's presence in the office, as well as the corresponding light conditions.

The first goal of the monitoring system is to identify whether there is user presence in the office or not and consequently to recommend user to switch off monitors and  lights when they are not necessary (e.g. when the user is away from the office or when the natural light is sufficient). The second goal of our monitoring system is to assist the user to keep a desired room temperature based on outdoor conditions, but also taking into account the user preferences and the identified user tendencies on air-condition usage.

\subsection{System setup}
In order to support the aforementioned case, a set of sensors and actuators has been deployed in two main areas, the office and the balcony in front of the office \footnote{An office at the Harokopio University of Athens has been used for testing.}. The aim was to capture not only indoor but outdoor conditions as well. For the indoor conditions, we are interested in capturing user presence, temperature, humidity and luminosity and power consumption, while the outdoor conditions monitored outside are temperature, humidity and luminosity.

For the purpose of demonstrating the efficiency of our proposed method we have deployed two types of sensor/actuator devices. The first type was commercial IoT devices from Sonoff (e.g. POW meters, Smart Switches, etc.) \footnote{\url{https://sonoff.itead.cc/en/}} and the second type was custom type of devices based on EPS8266 chip (NodeMCU, Wemos). The ESP8266 chip was selected based on the capability of its 2.4GHz wireless connectivity and low power consumption \footnote{\url{https://www.espressif.com/sites/default/files/documentation/esp8266-technical_reference_en.pdf}}. In both types of devices, the firmware was replaced by the TASMOTA firmware \footnote{\url{https://github.com/arendst/Sonoff-Tasmota/wiki}} using an FTDI USB programmer. The TASMOTA firmware is an open-source firmware that offers more control and better configuration on any device based on the ESP8266 chip.

In order to monitor consumption and in the same time being able to remotely control the devices, we used  two Sonoff POW devices, one for the monitors and the other for the A/C unit. The Sonoff POW devices can control (turn-on/off) the devices connected to them and at the same time, record the power consumption in real-time. 
In addition, we have installed an RF 433MHz wireless motion sensor to detect user presence in the office and a Sonoff Bridge device for the communication of any RF 433MHz module with the Home Assistant dashboard. Regarding the light control of the office, we have used a Sonoff Switch as a wireless switch to control switching on or off the lights when necessary.

Finally, we have installed a custom type of device using a NodeMCU chip, with a motion sensor (AM312 chip\footnote{\url{http://www.image.micros.com.pl/_dane_techniczne_auto/cz\%20am312.pdf}}) connected on it, a humidity/temperature sensor (DHT11\footnote{\url{https://www.mouser.com/ds/2/758/DHT11-Technical-Data-Sheet-Translated-Version-1143054.pdf}}), a light sensor (analog photoresistor) and an IR transmitter. The NodeMCU unit had the role of multi-sensor and at the same time with the IR transmitter had the capability to control the air conditioning unit through the infrared signal communication. 

\begin{figure}[h!]
\includegraphics[trim={1in 0.8in 1in 1in},clip,width=1.0\columnwidth]{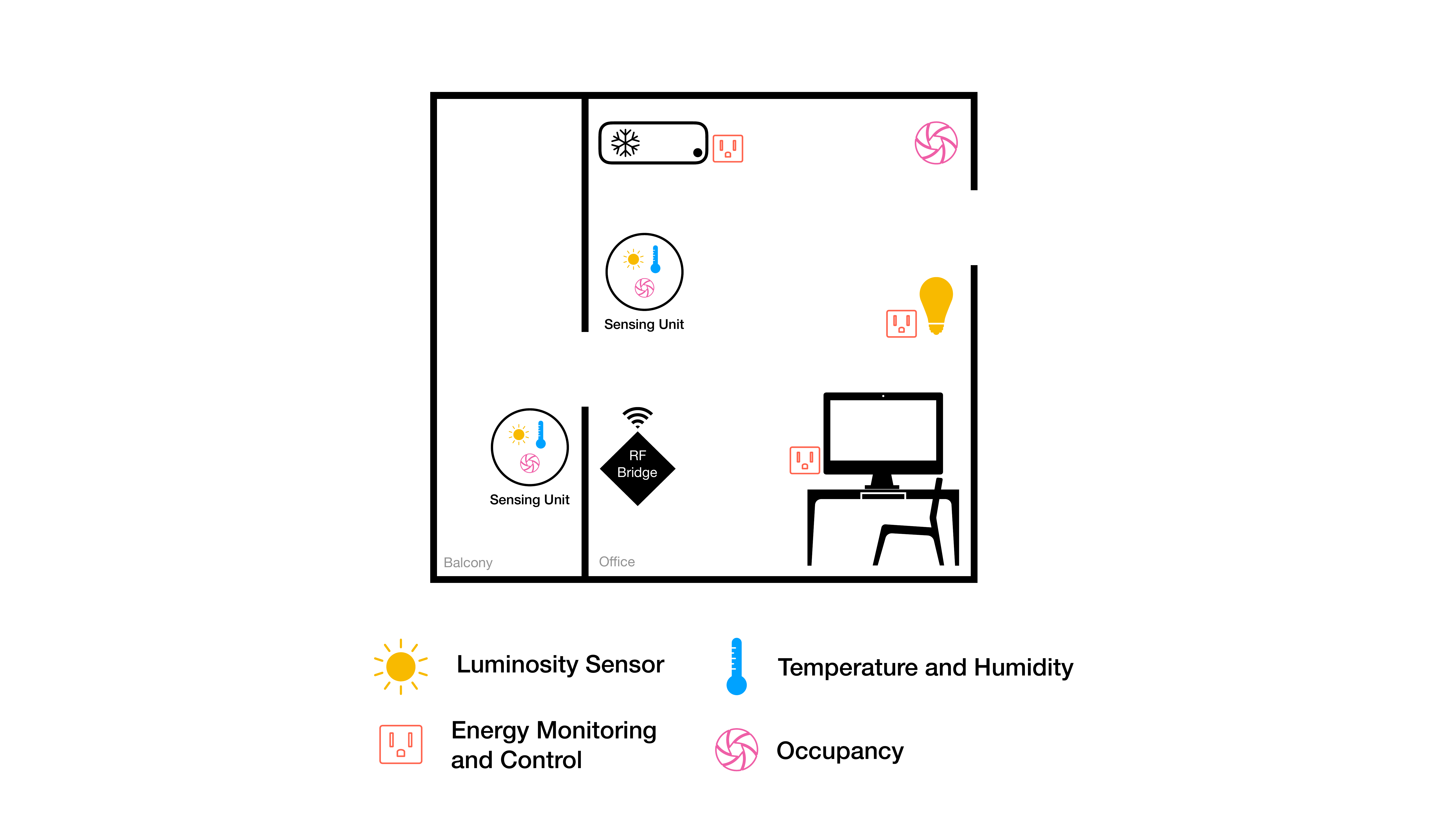}
\centering
\caption{The deployment of the sensors in our facilities.}
\label{fig:sensors-diagram}
\end{figure}

At the balcony area, we used a custom prototype of the device (NodeMCU) connected with a humidity/temperature sensor and a light sensor. All the custom prototype devices were enclosed in custom design 3D printed boxes for the protection of the circuit, the wiring and for the smooth integration with the building rules of aesthetics. Figure \ref{fig:sensors-diagram} illustrates the top view of the area that our pilot was deployed.

\subsection{Data management and integration}
\begin{figure}[h!]
\includegraphics[width=0.9\textwidth]{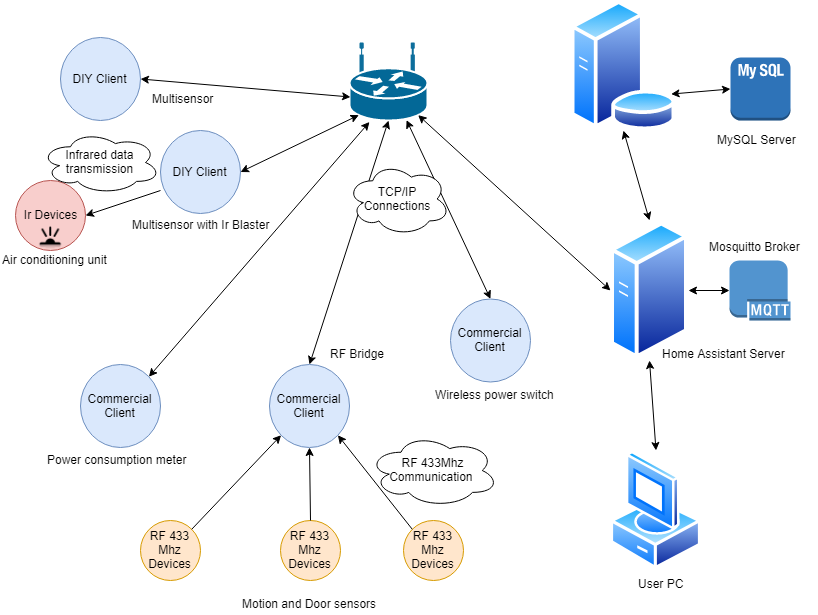}
\centering
\caption{Network connectivity of the components of the system.}
\label{fig:system-diagram}
\end{figure}

As depicted in Figure \ref{fig:system-diagram} all client modules are connected through TCP/IP via Ethernet or Wi-Fi connections with a Raspberry Pi device, which also hosts the Home Assistant server. The communication between the server and the clients was based on exchanging data using MQTT communication protocol, which is also hosted as a service in the Home Assistant server. The MQTT (Message Queuing Telemetry Transport) is an ISO standard (ISO/IEC PRF 20922) publish-subscribe-based messaging protocol and it was selected based on: \textit{i)} the requirement of minimizing network traffic, \textit{ii)} the ability to increase the security of the connection (SSL security) and \textit{iii)} the extra feature of different modes for Quality of Service (QoS).

Two different QoS modes have been employed depending on how crucial is the connection to a device. The switches use the QoS mode 2, in which the sender and receiver engage in a two-level handshake to ensure only one copy of the message is received. The measurements employ the QoS mode 0, in which the message is sent only once and the client and broker take no additional steps to acknowledge delivery. For the modules controlled via infrared (IR) (e.g. Air conditioning units, TVs, etc.) on the initial configuration, an extra Arduino board (e.g. Arduino Nano) with an IR receiver connected to it has been employed.

In order to control the A/C unit from the Home Assistant platform, it is necessary to decode all the remote control signals and map them to different commands in binary format. This was done using the Arduino board and an IR receiver to collect the transmitted binary codes from the remote control. These codes were later used in the Home Assistant User Interface (UI) for creating different automations for different application scenarios. Finally, a set of motion detection sensors, which are very common in home security alarm systems, that employ the RF 433MHz wireless communication protocol have been used. The main advantages of this type of devices are their low cost and their ease of configuration and installation \cite{qingyun2008433mhz}. In addition, they are battery operated and can be readily installed in all buildings. When a motion sensor is triggered, then a unique code is published and every device with the capability to receive RF 433MHz signals can ``capture" this code and based on that may trigger predefined procedures that record an  action (e.g. switch-on/off a light, enter or exit a room, etc.)
For controlling all the RF 433MHz signals,  an extra receiver has been used to collect  signals and transmit them to the server through the Wi-Fi connection. The messages are then used by the Home Assistant automations, as shown in Figure \ref{fig:automations}.

The data storage of the  system, is handled by a MySQL server, running on a PC in the same network, instead of the NoSQL option that comes as default with Home Assistant. This option allows better scalabilty of the system and moves part of the processing load from the Raspberry Pi to another device. It also allows to further expand the data analysis and processing and the monitoring system as a whole.

\subsection{User interface and automations}
In order to integrate the whole system in a user-friendly automation hub, we employed Home Assistant\footnote{https://www.home-assistant.io/}. It is an open source automation platform that puts local control and privacy first and is powered by a worldwide community of DIY enthusiasts. In our case, the host for running Home Assistant is a Raspberry Pi 3 Model B and serves as the server part of our system architecture.

More specifically, the Home Assistant platform offers a very comfortable means to track the collected data and at the same time offers a graphical representation of the data using custom graphs. The UI is accessible through a web browser and has an easily configured layout. The user can modify the layout and monitor multiple rooms and is able to control the connected devices using virtual switches. The state of every switch gets updated automatically if the state changes through the smart device or manually. In addition, the Home Assistant UI offers the users the ability to construct custom automations following the simple flow of:  ``TRIGGER $->$  CONDITION $->$ ACTION".

\begin{figure}[h!]
\includegraphics[width=0.9\textwidth]{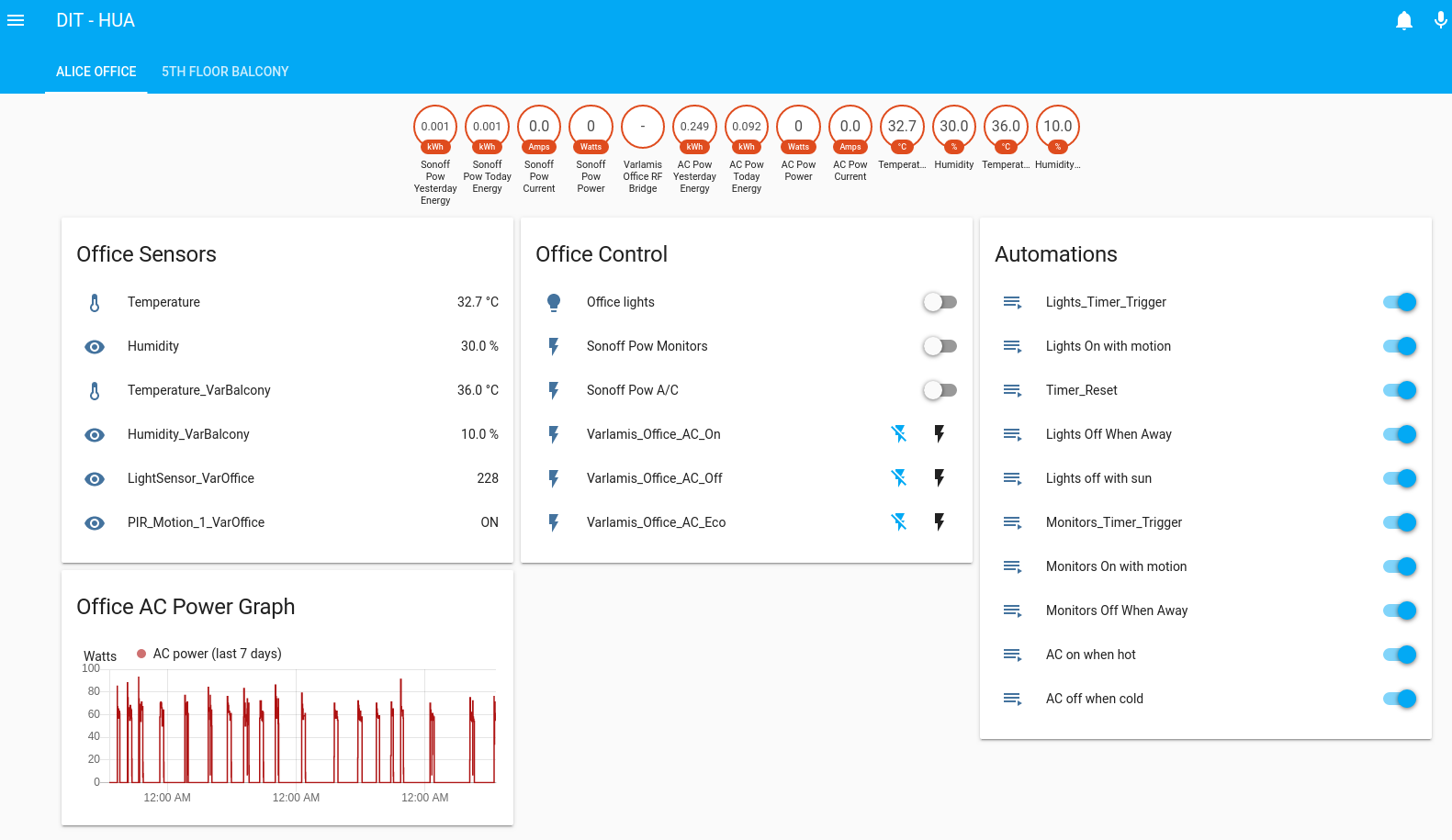}
\centering
\caption{The main layout of the Home Assistant UI, comprising sensors, switches, automations and analytics.}
\label{fig:home-assistant-ui}
\end{figure}

\begin{figure}[h!]
\includegraphics[width=0.55\textwidth]{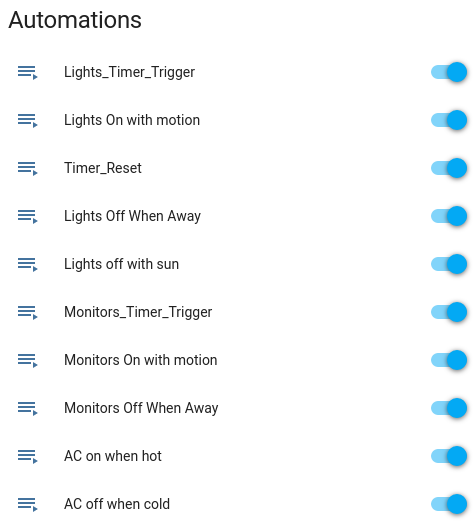}
\centering
\caption{The list of automations that implement the case scenario.}
\label{fig:automations}
\end{figure}

Figure \ref{fig:home-assistant-ui} demonstrates the Home Assistant's main interface for the case scenario. The layout contains information about sensors' current status, switches and analytics that provides user with instant monitoring of sensor logs or consumption of the office appliances that have been integrated in the monitoring system. At the right of Figure \ref{fig:home-assistant-ui}, there is the ``Automations" section, which is also depicted in Figure \ref{fig:automations}, and includes all the custom automations that control the office infrastructure (Actions) and implement energy saving and habit change scenarios (i.e. rules with \textbf{Triggers} and \textbf{Conditions}, which start the appropriate (\textbf{Action}) when met.).
The actions where not directly triggered, but using the Telegram cloud-based messaging API\footnote{https://core.telegram.org/} that is supported by Home Assistant, a recommendation is sent to the user. When the user accepts the recommendation, then the Action happens.

\section{Results and Discussion}\label{sec:results}

In order to control the usage of office lights, we created several triggers (Automations) that either turn on the lights, when somebody is in the office and the light levels from the sunlight are low, or turn off the lights after 15 minutes of inactivity in the room, based mainly on a rule-based logic. Since the experiment was performed in an office, the expected office usage times are from 8 a.m. to approximately 8 p.m., only on weekdays and with many breaks, since the office is not always occupied.

\begin{figure}[htb!]
\centering 
\includegraphics[width=1\textwidth]{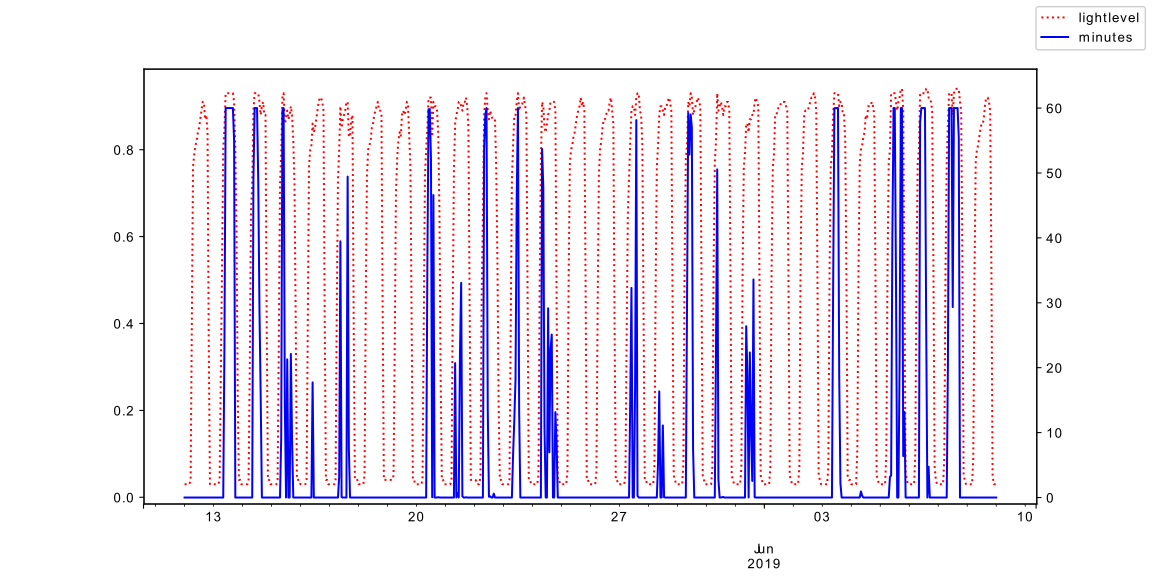}
\caption{Light levels of the office for a period of 4 weeks and the respective lights usage time per hour. The left vertical axis measures the light level (red line) of the office in a scale from 0 (dark) to 1 (light). The right vertical axis measures the hourly usage of lights per hour (blue line) with a maximum of 60 minutes.}
\label{week1_4_lights_luminocity}
\end{figure}

Figure \ref{week1_4_lights_luminocity} shows the light levels of the office for a period of 4 weeks (weeks 2 and 3 with the automations enabled and weeks 1 and 4 without automations) and the respective lights' usage time per hour (in minutes). We can see that using the automations, the office light levels (scaled between 0 for dark and 1 for maximum light) followed almost the same pattern, whereas the usage time of the office lights is limited during the day (when the office is occupied), with the few peaks to correspond to a continuous user presence in the office. Even in this case, the maximum usage time per hour (i.e. 60 minutes) was reached only a few times during weeks 2 and 3, mainly because of the plenty natural sunlight that triggered the turn off notifications. It is also obvious from the plots that the consumption during the night or the weekend is zero. From a power consumption point of view, it is worth noting that when using the automations, the average lights’ usage time was 2.98 and 2.04 Hours (for week 2 and 3 respectively), whereas during the period that no automations were used it was 3.78 and 5.03 hours per day (for weeks 1 and 4 respectively).

The impact of our monitoring system is highlighted during weeks 1 and 4 when all the automations have been deactivated leading to an increase of more than 75\% of the average daily usage time for weeks 2 and 3.

\begin{figure}[htb!]
\centering 
\includegraphics[width=0.9\linewidth]{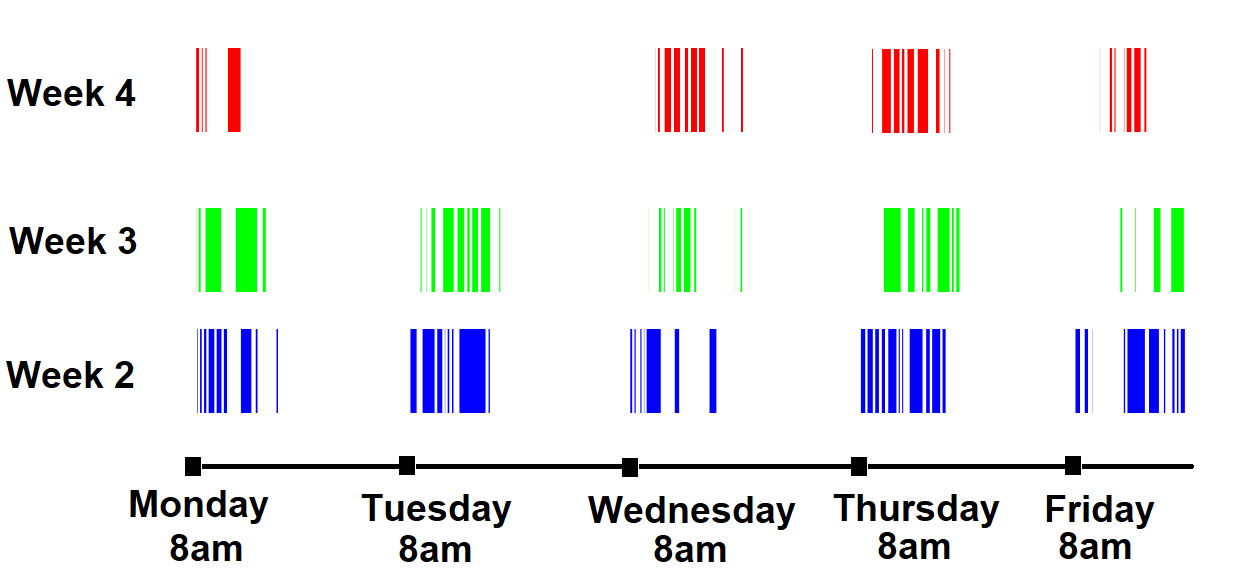}
\caption{The time slots during which user movement was detected at the office.}
\label{inoffice_timeplots}
\end{figure}

\begin{figure}[htb!]
\centering 
\includegraphics[trim={0in 0.2in 0in 0in},clip,width=1.0\linewidth]{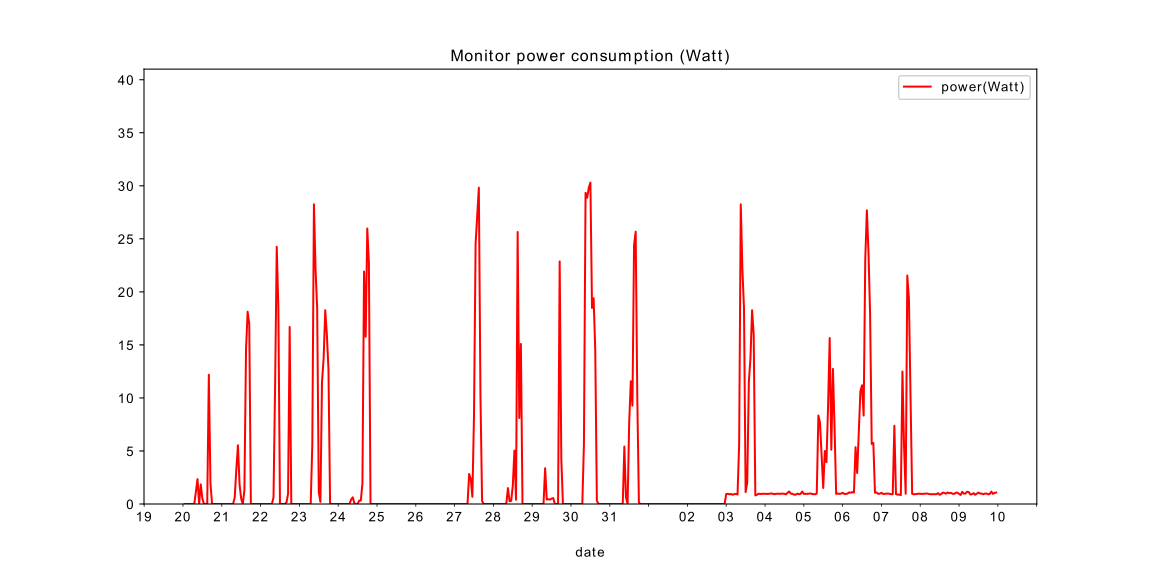}
\caption{Power consumption in Watts for the two office monitors for weeks 2 to 4. During week 4 no automation have been used apart from the stand by and sleep mode options of the monitors.}
\label{monitor-kwh}
\end{figure}

Evaluating the process of automatically switching off the monitors when Alice was out of the office for more than 15 minutes, in Figure \ref{inoffice_timeplots}, we plot the periods during which Alice was at the office, whereas Figure \ref{monitor-kwh} contains the respective power consumption during these weeks from the two office monitors. The monitors’ scenario ran for weeks 2 to 4, with automations turned on in week 2, with partial use of the automations in week 3 and without any automations in week 4.

As Figure \ref{inoffice_timeplots} implies, the office usage patterns were quite similar across weeks and most importantly they contain a lot of idle periods in which the office, and thus the monitors, are not used. 
Taking advantage of the automations that was used, Alice managed to reduce monitor consumption especially in weeks 2 and 3.

Based on the consumption values of the two monitors during standby ($\approx$ 0.5 Watt each) and normal operation ($\approx$ 38 Watt each), it is possible to draw some useful conclusions. First, the sleep/standby consumption of the two monitors adds at least 168 Watts to this sum (i.e. 24 hours $\times$ 7 days $\times$ 0.5 Watt $\times$ 2 monitors), which justifies this 42\% increase compared to week 2 and the 21\% increase compared to week 3. Second, given that the user was at the office for 5 days, the actual monitor usage time per day for the three weeks is almost the same and ranges from 1 to 1.3 hours. Although the savings for one user may not be impressive (i.e. the yearly savings will be a few dollars), projecting them to an office building or to a larger scale can be useful. It is also important to highlight that using similar automations for more devices, we shape user habits towards energy efficiency.

\section{Conclusions and next steps}
\label{sec:conclusions}

Addressing the problem of engaging users in adopting more sustainable energy usage tactics, we identify that the users' everyday energy-related behavior is driven by their needs and desires. Targeting the user's repetitive common energy consumption tasks we can create triggers that will automatically act instead of the user to lead to more efficient patterns of energy consumption.

In this work we extend our previous work on reshaping user consumption habits towards efficiency and present an implementation of a real-case scenario of our monitoring system in our facilities as a proof of concept of automating energy consumption management task. Current results show that leveraging the benefits of an open-source platform like Home Assistant and a setup of smart sensors, smart plugs and a consumption reporting mechanism we are able to facilitate the transition of the user to new consumption habits (like efficiently operating the office lights, A/C, etc.).

We are currently in the process of expanding our architecture in all of the areas of our department, that will not only allow us to efficiently use energy source in all of the department's infrastructures but will verify our system's benefits in a large-scale setup.

\section*{Acknowledgements}

This paper was made possible by National Priorities Research Program (NPRP) grant No. 10-0130-170288 from the Qatar National Research Fund (a member of Qatar Foundation). The statements made herein are solely the responsibility of the authors.


\end{document}